\documentclass[twocolumn,floatfix,superscriptaddress,amsmath,showpacs,showkeys,aps,prb]{revtex4}
\usepackage[utf8]{inputenc}
\usepackage[final]{graphicx}
\usepackage{t1enc}
\usepackage{bm}
\usepackage{times}

\begin{document}

\bibliographystyle{apsrev}

\title{T=0 phase diagram and nature of domains in ultrathin ferromagnetic films with perpendicular anisotropy}

\author{Santiago A. Pigh\'{\i}n}
\email{pighin@famaf.unc.edu.ar}
\affiliation{Facultad de Matemática, Astronomía y Física, Universidad Nacional de Córdoba and Instituto de Física Enrique Gaviola  (IFEG-CONICET)\\
Ciudad Universitaria, 5000 Córdoba, Argentina}

\author{Orlando V. Billoni}
\email{billoni@famaf.unc.edu.ar}
\affiliation{Facultad de Matemática, Astronomía y Física, Universidad Nacional de Córdoba and Instituto de Física Enrique Gaviola  (IFEG-CONICET)\\
Ciudad Universitaria, 5000 Córdoba, Argentina}
\author{Daniel A. Stariolo}
\email{stariolo@if.ufrgs.br}
\affiliation{Departamento de Física,
Universidade Federal do Rio Grande do Sul
and
National Institute of Science and Technology for Complex Systems\\
CP 15051, 91501-970 Porto Alegre, RS, Brazil}
\altaffiliation{Research Associate of the Abdus Salam International Centre for
Theoretical Physics, Trieste, Italy}
\author{Sergio A. Cannas}
\email{cannas@famaf.unc.edu.ar}
\affiliation{Facultad de Matemática, Astronomía y Física, Universidad Nacional de Córdoba and Instituto de Física Enrique Gaviola  (IFEG-CONICET)\\
Ciudad Universitaria, 5000 Córdoba, Argentina}

\date{\today}

\begin{abstract}

We present the complete zero temperature phase diagram of a model for ultrathin films with
perpendicular anisotropy. The whole parameter space of relevant coupling constants is
studied in first order anisotropy approximation. Because the ground state is known to
be formed by perpendicular stripes separated by Bloch walls, a standard variational
approach is used, complemented with specially designed Monte Carlo simulations.
We can distinguish four regimes according to the different
nature of striped domains: a high anisotropy Ising regime with sharp domain walls, a
saturated stripe regime with thicker walls inside which an in-plane component of the
magnetization develops, a narrow canted-like regime, characterized by a sinusoidal variation of both the in-plane and the out of plane magnetization components, which upon further decrease of the anisotropy leads to an
in-plane ferromagnetic state via a spin reorientation transition (SRT). The nature of
domains and walls are described in some detail together with the variation of domain
width with anisotropy, for any value of exchange and dipolar interactions. Our results,
although strictly valid at $T=0$, can be valuable for interpreting data on the evolution
of domain width at finite temperature, a still largely open problem.
\end{abstract}

\pacs{75.40.Gb, 75.40.Mg, 75.10.Hk}
\keywords{ultrathin magnetic films, Heisenberg  model, stripe width}

\maketitle
\section{Introduction}

 The magnetic phases of ferromagnetic
thin films with perpendicular anisotropy have been the subject of
intense experimental~\cite{AlStBi1990,
VaStMaPiPoPe2000,PoVaPe2003,WuWoSc2004,WoWuCh2005,VaBlLa2008,FrStMeOe2008},
theoretical~\cite{CaLe1971,YaGy1988,
CzVi1989,PePo1990,AbKaPoSa1995,Po1998,ZaStMa2007} and
numerical~\cite{DeMaWh2000,VeOeKi2002,CaStTa2004,PiCa2007,NiSt2007,CaBiPiCaStTa2008}
work in the last 20 years.
Magnetic order in ultrathin ferromagnetic films is very complex due to the competition
between several different energy contributions, the most prominent being exchange and
dipolar interactions, together with a strong influence of shape and
magnetocrystalline anisotropies of the sample. These in turn are very susceptible to
the  growth conditions of the films~\cite{Portmann2006,VaBlLa2008}.

A widely used model
that contains the main ingredients of ultrathin film magnetism is the 2D dimensionless
Heisenberg Hamiltonian:

\begin{widetext}
\begin{equation}
{\cal H} = -\delta \sum_{<i,j>} \vec{S}_i \cdot \vec{S}_j +
\sum_{(i,j)} \left[ \frac{\vec{S}_i \cdot \vec{S}_j }{r_{ij}^3} - 3 \,
\frac{(\vec{S}_i \cdot \vec{r}_{ij}) \; (\vec{S}_j \cdot \vec{r}_{ij})}{r_{ij}^5} \right]
- \eta \sum_{i} (S_i^z)^2
\label{hamiltoniano}
\end{equation}
\end{widetext}
where $\vec{S}_i$ are classical unit vectors, the exchange and anisotropy constants
are normalized relative to the dipolar coupling
constant ($\delta \equiv J/\Omega, \eta \equiv K/\Omega$), $<i,j>$
stands for a sum over nearest neighbors pairs of sites in a  square
lattice, $(i,j)$ stands for a sum over {\it all distinct} pairs and
$r_{ij}\equiv |\vec{r}_i - \vec{r}_j|$ is the distance between spins
$i$ and $j$.

At low temperatures and strong enough perpendicular anisotropy, the presence of
a striped phase (i.e., a modulated pattern of local
perpendicular magnetization with a well defined stripe width $h$) is well
established and is the ground state of the system~\cite{YaGy1988,CzVi1989,KaGe1993}.
In the limit of strong uniaxial anisotropy
domain walls are sharp and the energy cost for deforming or moving a domain wall is
large. Nevertheless, even when the mechanism by which the width of domains adjusts is not well understood, the stripe width varies with the effective anisotropy. When the thickness of the films (or the temperature) grows, the
effective perpendicular anisotropy is reduced in films of a few monolayers, and magnetostatic
energy becomes
important, inducing the magnetization to develop an in-plane component. Domains become
narrower, walls become
wider and are of Bloch type at low temperatures\cite{WuWoSc2004,Be1998},
until the system goes through
a Spin Reorientation Transition (SRT) when anisotropy and dipolar energies cancel
~\cite{MoUs1995,MiKi1996,ZdBa2003,WoWuCh2005}.
Around the SRT line a canted state may develop, where the magnetization presents a
finite in-plane
component together with the perpendicular modulation. The extension of
the canted state in parameter space strongly depends on the nature of the relevant anisotropies.
 For some systems, like Co/Au(111) , it seems necessary to go beyond
the first order anisotropy approximation of the model (\ref{hamiltoniano}). A
second order anisotropy energy is responsible for a canted state in a large portion
of the phase diagram~\cite{MiKi1996,VeOeKi2002,ZaStMa2007}. For other systems,
like Fe/Cu(001),  the first order anisotropy seems to be enough to describe the relevant physics\cite{GaRi1986, PlAnFaBa1999}.
In this case, as described by the model (\ref{hamiltoniano}), the canted state is restricted to a narrow region of
parameter space around the SRT, as expected from general
considerations~\cite{VeOeKi2002}, and reported in simulations at
finite temperature~\cite{WhMaDe2008}.
Finally, when the dipolar anisotropy exceeds the magnetocrystalline one, the system
enters an in-plane ferromagnetic state.

In this work we extend previous calculations~\cite{YaGy1988,MaDeWh1998,Po1998} and compute the
complete phase diagram
 in the $(\delta,\eta)$ space of Hamiltonian
(\ref{hamiltoniano}), at $T=0$. We also improve upon previous results by considering
different kinds of domain walls (sinusoidal, hyperbolic tangent, sharp walls), as
appropriate for each regime in parameter space.
We consider only straight domains, (domains in which the spin orientation
can be modulated along the $x$ direction but is constant in the
perpendicular direction $y$) separated by Bloch walls, i.e., walls
in which the magnetization stays inside the $yz$ plane. The local magnetization
vector inside the domains may be tilted at
an angle $\theta$ with respect to the plane normal (z axis).
Within these assumptions, we
obtain the complete phase diagram, the variation of the angle $\theta$ and
the behavior of the width of domains and walls in the whole parameter space $(\delta,\eta)$.
This allows, e.g., to obtain the crossover between Heisenberg (extended walls) and Ising
(sharp walls) regimes. We verify that domain width adjustment with varying anisotropy
is only possible in the Heisenberg regime, domain width being fixed in the Ising regime
for any value of the parameters $(\delta,\eta)$.
Analytical calculations are complemented with  Monte Carlo simulations specially
designed for the present purposes, as explained in the appendix.

\section{Zero temperature phase diagram}
\label{phasediagram}

We consider a square lattice with $N=L\times L$ sites,
characterized by the integer indexes $(x,y)$, where $-L/2 \leq x \leq
L/2$ and $-L/2 \leq y \leq L/2$, in the limit $L\to\infty$. Hence, the
index $i$ in Eq.(\ref{hamiltoniano}) denotes a pair of coordinates
$(x,y)$. We  consider only
uniformly magnetized solutions along every vertical line of sites,
i.e. $\vec{S}_{(x,y)}=\vec{M}(x)$, $\forall y$ and allow only Bloch
walls between domains of perpendicular magnetization,
 i.e. $M^x(x)=0$ $\forall x$.
Yafet and Gyrogy (YG) showed that for these types of  spin configurations the energy per spin
can be mapped onto the  energy of a one dimensional XY model~\cite{YaGy1988}. The energy
difference between an arbitrary magnetization profile $\vec{M}(x)$ and a uniformly in--plane
magnetized state is then given by:

\begin{widetext}
\begin{equation}
e\left[\vec{M}(x)\right]= (\delta-2c_2) - \frac{\delta'}{L}
\sum_x \vec{M}(x). \vec{M}(x+1) +
 \frac{1}{L} \sum_{x,x'}
\frac{M^z(x)\, M^z(x')}{|x-x'|^2} - \frac{\kappa'}{L} \sum_x
\left[M^z(x) \right]^2 + C
\label{energia1}
\end{equation}
\end{widetext}

\noindent where $\delta'=\delta-2\, c_1$, $\kappa'=\eta-3\, g$, $c_1=0.01243\ldots$, $c_2=0.07276\ldots$, $g=1.202057\ldots$ and

\begin{equation}\label{correction}
    C \equiv C\left[M^y(x)\right]= 2(c_2-c_1)\frac{1}{L} \sum_x M^y(x)\, M^y(x+1)
\end{equation}

\noindent Although small, this correction term makes a non negligible
contribution when the domain walls are of the same order of the
lattice constant. This happens for small values of $\delta$ ($\delta <
5$), where both the stripe and wall widths are of the order of a few
lattice spacings. For larger values of $\delta$ it is reasonable to assume a smooth magnetization profile\cite{YaGy1988} $M^y(x+1)\approx M^y(x)$, so that the correction (\ref{correction})
 can be absorbed into the anisotropy term in
Eq.(\ref{energia1}),  replacing $\kappa'\to\kappa= \eta-3\, g+2(c_2-c_1)$.

Now consider a stripe-like periodic structure of the magnetization profile  with period $2h$, $ M^z(x+h)= -M^z(x)$. Using a Fourier expansion:

\begin{equation}\label{Fourier-stripes}
    M^z(x) = M_0 \sum_{m=1,3,\ldots} b_m \; \cos{ \left(\frac{m\pi\, x}{h} \right)},
\end{equation}

\noindent the energy (\ref{energia1}) can be written as\cite{YaGy1988}

\begin{widetext}
\begin{equation}
e\left[\vec{M};\delta,\eta \right]= (\delta-2c_2) - \delta'
\frac{1}{L} \sum_x \cos{\left[ \phi(x)-\phi(x+1)\right]} +  M_0^2
\sum_{m=1,3,\ldots} b_m^2  D_m(h)- \frac{\kappa'\, M_0^2}{2}
\sum_{m=1,3,\ldots} b_m^2 + C
\label{energiabm}
\end{equation}
\end{widetext}

\noindent where $\phi(x)$ is the angle between $\vec{M}(x)$ and the $z$ axis and

\begin{equation}\label{Dm}
    D_m(h) \equiv \sum_{u=1}^\infty \frac{\cos{(m\pi\, u/h)}}{u^2}= \frac{\pi^2}{6} - \frac{\pi^2m}{2h} + \left( \frac{\pi m}{2h}\right)^2.
\end{equation}

Now we  look for the minimum of Eq.(\ref{energiabm}) for
different values of $\delta,\eta$. We propose  different  striped
magnetization profiles $M^z(x)$ and compare the energies obtained by
minimizing Eq.(\ref{energiabm}) for each profile with respect to variational parameters.

We first consider a profile as proposed by YG, that is constant $|M^z(x)| =M_0$ inside
the stripes with a sinusoidal variation inside the walls between
stripes (see Fig.1 in Ref.~\cite{YaGy1988}). This will be called ``sinusoidal
wall profile approximation'' (SWP). In order to allow for canted
profiles, we take $M_0= \cos \theta$, where $\theta$ is the canting
angle, i.e. we define it as the minimum angle of the local
magnetization with respect to the $z$ axis. In Ref.~\cite{YaGy1988}
this variational problem was  solved for $M_0=1$ in the continuum
limit, i.e. when $h\gg 1$ and the wall width $w\gg1$, so that the
profile can be considered a smooth function of $x$. While this
approximation is expected to work well for large enough values of
$\delta$, it breaks down for relatively small values of it, where the
discrete character of the lattice has to be taken into
account. However, the variational problem for that range of values of
$\delta$ can be solved exactly (although numerically) by minimizing
Eq.(\ref{energiabm}) with respect to the {\it integer} variational
parameters $h$ and $w$ and continuous parameter $\theta$. In other
words, for every pair of values $(\delta,\eta)$ we evaluate the energy
Eq.(\ref{energiabm}) for the sinusoidal profile with different
combinations of $h=1,2,\ldots$ and $w=1,2,\ldots$ within a limited
set. For every pair of values $h,w$,we look for the value of $\theta$
that minimizes the energy with a resolution $\Delta \theta = 0.01$ and
compare all those energies.
This  calculation is feasible for values  up to $\delta=10$, for which the maximum value of $h$ (bounded by the stripe width in the $\eta\to \infty$ limit) remains relatively small (smaller than $h=140$). Some results for $\delta=15$ close to the SRT were also obtained. All the results of this calculation are compared against Monte Carlo (MC)  simulations. Details of the MC method used are given in Appendix \ref{MCmethod}. Through these calculations we obtain a zero temperature phase diagram for low values of $\delta$.

Before presenting the results, it is important to introduce some
notations and definitions of different types of solutions. We
distinguish between four types of solutions. If the minimum energy
solution corresponds to $w=1$ and $\theta=0$ (within the resolution
$\Delta \theta$), we call this a {\it Striped Ising Profile} (SIP),
i.e. a square wave like profile. If $\theta=0$ but $w > 1$, we call
this a {\it Saturated State} (SS). These states only show a finite in-plane
component of the magnetization inside the walls. If $ 0 < \theta < \pi/2$ the solution is
a canted-like state. Finally, if $\theta = \pi/2$ ($M_0=0$) we have a {\it Planar Ferromagnet} (PF).

\begin{figure}
\begin{center}
\includegraphics[scale=0.27]{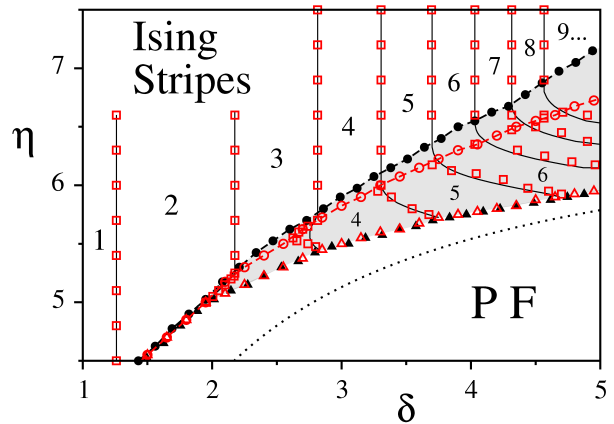}
\caption{\label{pd1} (Color online) Zero temperature phase diagram for
  small values of $\delta$. Black filled symbols and black solid
  lines: MC simulations. Open red symbols: SWP approximation. Squares
  and continuous black lines correspond to transition lines between
  striped states of different width. The shaded region corresponds to
  a canted-like state ($0 < \theta < \pi/2$).
Triangles are transition lines between  Planar Ferromagnet
  ($\theta=\pi/2$) and SMCP States (Spin Reorientation Transition
  line). Circles  mark transitions between the canted-like and the Striped
  Ising state ($\theta=0$ and $w=1$). Notice the excellent agreement
  between the MC and SWP calculations close to the SRT, while the SWP
  approach underestimates the transition line between the canted-like and
  Ising Stripes states. The dotted line corresponds to the continuum
  approximation of YG for the SRT (Eq.(\ref{etaSRT})).}
  \end{center}
\end{figure}

The zero temperature phase diagram for small values of $\delta$
($\delta \leq 5$) is shown in Fig.\ref{pd1}. For relatively large
values of $\eta$ the minimum energy configuration is always the Ising
one (SIP), with a stripe width independent of $\eta$. For small values
of $\eta$ the minimum energy configuration is the PF, with a spin
reorientation transition line (SRT), either to the Ising state for $h
<3$ ($\delta\sim 2)$) or to a canted-like one for $h\geq 3$ ($\delta >
2$). No Saturated State configurations are observed for $\delta<6$.

Inside the canted region, a strong stripe width variation with the
anisotropy is observed at constant $\delta$ . Note that the vertical
lines that separate Ising striped states with consecutive values of
$h$ bend inside the canted region and become almost horizontal as $\delta$
increases. Hence, the exponential increase of $h$ with $\delta$ in the
Ising region (vertical lines) changes to an exponential increase with
$\eta$ inside the canted region (curved lines on the right of Fig.\ref{pd1}).
It is important to note here that the canted region in this system
corresponds almost everywhere (except close to the crossover to an Ising striped state) to a regime in which the stripe width and walls are of the
almost equal, which means a pure sinusoidal magnetization profile. In this sense it has a different character than
the ``true'' canted phases obtained in systems with non-zero higher order
anisotropies~\cite{VeOeKi2002,ZaStMa2007}, where well defined domains show a finite in-plane magnetization component. In the present case the canted like states are characterized by a sinusoidal variation (with wave length $2h$) in both the in-plane and the out of plane magnetization components, without well defined  domains (see an example in Fig.\ref{pd2}). Hence, there are not truly ``stripes'', but a sinusoidal modulated state or Single Mode Canted Profile (SMCP).

We also find an excellent agreement between the sinusoidal wall
 approximation (or SMCP) and the MC results, except close to the
transition between the Ising and the canted-like states. Such disagreement
is due to the fact that the actual wall is not well described by a
sinusoidal profile far away of the SRT line, as will be shown later.

For large enough values of $\delta$ the variational problem for the SWP can be solved in a continuum approximation introduced by YG\cite{YaGy1988}, giving a set of coupled non-linear equations for the stripe width $h$, the ratio  between the stripe and the wall widths $\Delta= w/h$ and the canted angle $\theta$. In the limit $\Delta\to 1$  those equations can be solved analytically  predicting a SRT at the line\cite{YaGy1988}

\begin{equation}\label{etaSRT}
    \eta_{SRT}(\delta)= a- \frac{\pi^2}{2\delta}
\end{equation}

\noindent with $a=\pi^2/3+3g-2(c_2-c_1)$. The line Eq.(\ref{etaSRT}) is also depicted in Fig.\ref{pd1}. Notice the disagreement between the continuum approximation and the exact one for $\delta \leq 5$. This discrepancy becomes smaller than $1\%$ only for $\delta > 7$.

For arbitrary values of $\eta$ and $\delta$ the equations for $h$, $\Delta$ and $\theta$  can be solved numerically. From the numerical solutions we found that the range of values of the anisotropy $\eta$ for which the canted angle is appreciable different from zero within this approximation is strongly depressed as $\delta$ increases. For values $\delta\sim 100$ the canted-like configurations almost disappear, except very close to the reorientation line, as already  reported by Politi\cite{Po1998}.

 \begin{figure}
 \begin{center}
    \includegraphics[scale=0.27,angle=-90]{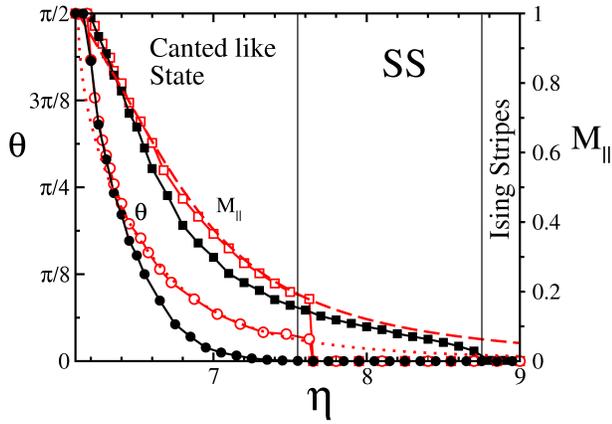}
    \caption{\label{mxSvsEta} (Color online) Canted angle (circles) and in-plane
      magnetization (squares) as a function of $\eta$ for
      $\delta=7.5$. Filled black symbols correspond to MC
      calculation. Open red symbols  corresponds to the discrete SWP
      approximation, while the red doted and dashed lines correspond
      to the continuum (YG) approximation of the SWP. Continuous black and red lines are only a guide to the eye.}
  \end{center}
  \end{figure}

Indeed, from our MC simulations, we observe that the range of values
for which  canted-like states have the minimum energy gradually shrinks as
$\delta$ increases, being replaced by a saturated state for values of
$\eta$ above certain threshold. This can be observed in
Fig.\ref{mxSvsEta}, where we show the behavior of the canted angle and
the in--plane magnetization component $M_{||}=(1/L)\sum_x M^y(x)$ as a
function of $\eta$ for $\delta=7.5$. The Monte Carlo data shows the
existence of a wide range of values of $\eta$ for which the canted
angle is zero while $M_{||}\neq 0$, meaning that the non null
in--plane components are concentrated inside the walls. In other
words, in that region we have a saturated state with thick walls $w
>1$. Notice also that the SWP approach completely fails to describe
those states. Moreover, we observe from our MC simulations that the
SWP cease to be the minimum energy solution for values of $\eta$
relatively close to the SRT,  before the saturated state sets up (see Fig.\ref{mxSvsEta}). This effect becomes more marked as $\delta$ increases.

 The departure of the magnetization profile from
 the SWP for large values of $\eta$ and $\delta$  is expected
 from micromagnetic theory, which  in that limit predicts that the
 wall structure will be dominated by the interplay between anisotropy and exchange, leading to an hyperbolic tangent shape of the wall\cite{Be1998}.  Hence we considered a periodic magnetization profile with hyperbolic tangent walls (HWP) defined, for a wall centered at $x=0$, by

\begin{equation}\label{hyperbolicprofile}
    M^z(x)= M_0\, \tanh{\left(\frac{x}{l_w} \right)} \;\;\;\; \mbox{for} \;\;\;\; -h/2 \leq x\leq h/2,
\end{equation}

\noindent  where $M_0=\cos{\theta}$ as before. In the large $\delta$ limit, assuming a smooth profile  $h \gg 1$ and $l_w\gg 1$, the anisotropy energy can be expressed as:

\begin{equation}
   e_{an} \approx - \kappa\,M_0^2 \left[ 1-\frac{2\,l_w}{h}\tanh{\left( \frac{h}{2\, l_w} \right)}\right]. \label{anitanh}
\end{equation}

\noindent The exchange energy can be obtained in a similar way:

\begin{widetext}
\begin{equation}
    e_{exc}=-\delta \left[ 1 -  \frac{l_w}{h} \left( \frac{M_0^2-1}{M_0} \tanh^{-1}\left(M_0 \tanh \left( \frac{h}{2l_w} \right)\right) \\
+ \tanh \left( \frac{h}{2l_w} \right) \right) \right]. \label{eexchHWP2}
\end{equation}
\end{widetext}

The dipolar energy can be calculated using Eq.(\ref{Dm}). The Fourier  coefficients for the profile (\ref{hyperbolicprofile}) can be computed using the approximation

\begin{equation}
    \tanh(x)\approx \left\{\begin{array}{ll}
             x(1-\frac{x^2}{3}) & if \;\; 0 \leq x \leq \frac{1}{2} \\
             (1-e^{-2x})^2(1+e^{-4x}) & if \;\; \frac{1}{2} \leq x
           \end{array} \right.\label{perfilTanh}
\end{equation}

\noindent This leads to an
expression for the total energy as a function of the variational
parameters $h$, $\theta$ and $l_w$ that can be minimized
numerically. Comparing the minimum energy solution for the SWP and the
HWP we obtain the crossover line between sinusoidal and hyperbolic
wall structure shown in Fig.\ref{pd2} (dashed line). Above that line
the HWP has always less energy than the SWP. We also calculated the
transition line between the canted-like and the saturated states by setting
the condition $\theta = 0.01$, to be consistent with the criterium
used in the MC calculations. The results are shown in Fig.\ref{pd2}
together with the SRT line Eq.(\ref{etaSRT}), and compared with MC
calculations up to $\delta=15$. The excellent agreement with the MC
results gives support to the analytic approximations. Only between the dashed and the continuous lines we found truly canted states (i.e., states with $h \ll w$), although they parallel component is rather small ($\theta < 0.1$). In this sense, the region enclosed by both lines marks a crossover between the SMCP and the Saturated states: as $\eta$ increases domains emerge gradually, the walls change from sinusoidal to hyperbolic shape and the canted angle goes to zero.

\begin{figure}
\includegraphics[scale=0.27]{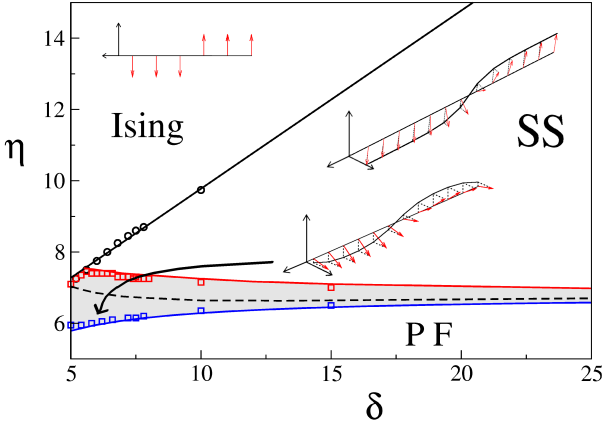}
\caption{\label{pd2} (Color online) Zero temperature phase diagram for
  large values of $\delta$. The shaded region corresponds to the canted-like states. Symbols correspond to MC simulations and lines to theoretical results. The dashed line correspond to the crossover between sinusoidal and hyperbolic wall structure. The lower line (blue) corresponds to Eq.(\ref{etaSRT}). The middle line (red) is obtained from the HWP minimum energy solution with $\theta = 0.01$. The upper line (black) corresponds to Eq.(\ref{etatanhYG}). Typical magnetization profiles obtained by MC are shown for every region of the phase diagram.}
\end{figure}

For large values of $\eta$ the exponential increase of $h$ makes it
cumbersome to apply the previous approximation
for the calculation of the dipolar energy. Instead of that, we can use
the following heuristic argument to obtain a reasonable
approximation. The main error introduced by the SWP approach is in the
exchange and anisotropy contributions to the energy. Since the main contribution to  the dipolar energy is given by the interaction between domains, we can assume that the dipolar contribution of the wall is relatively independent of its shape. Hence, we approximate the dipolar contribution by the SWP expression obtained by YG \cite{YaGy1988}
taking $w=f\, l_w$ ($f$ is a fitting parameter of order one to be fixed later) in
the limit $\Delta \ll 1$ ($l_w/h \ll1$). We compare the energy obtained within this approximation  with that obtained using the Eq.(\ref{perfilTanh}) for different values of the system parameters. We verified that the error made by taking $f=4$ is always smaller than $1\%$ for $h/l_w \geq 20$. We also observe that the best agreement with the MC results is obtained for $f=4$. Assuming then $M_0=1$,
the total energy per spin (relative to the parallel magnetized state)  for the HWP can then be approached by


\begin{equation}\label{energyHWP}
    e_{HWP}= \gamma +\frac{\delta/l_w-2\,l_w\,\gamma}{h}- \frac{4}{h}\, \ln \left( \frac{3\, \pi h}{10\,l_w} \right)
\end{equation}

\noindent with $\gamma=\pi^2/3 -\kappa$. Minimizing Eq.(\ref{energyHWP}) with respect to the variational
parameters $h$ and $l_w$ leads to:

\begin{equation}\label{htanhYG}
    h =  \frac{10}{3 \pi}\, l_w \exp \left[\frac{\delta}{2 l_w}\right],
\end{equation}

\noindent with

\begin{equation}\label{lwtanhYG}
    l_w  =  \frac{ \delta }{2 + \sqrt{ 4 + 2 (\kappa  -  \pi^2/3)\delta }},
\end{equation}

\noindent in agreement with a  derivation made by Politi\cite{Po1998}.

With the previous calculation we can also estimate the transition line between the saturated and the Ising Striped state. In the large $h$ limit the energy for a SIP, i.e. for

\begin{equation}\label{phi-ising}
    \phi(x)=\left\{\begin{array}{cc}
                    0 & \mbox{if} \;\; 0\leq x\leq h/2 \\
                    \pi & \mbox{if} \;\; h/2< x\leq h
                  \end{array}
    \right.
\end{equation}

\noindent the energy can be easily calculated from Eq.(\ref{energiabm}). The Fourier coefficients are:

\begin{equation}\label{bm-ising}
    b_m=(-1)^{(m-1)/2} \frac{4}{\pi m}.
\end{equation}

\noindent Using  Eq.(\ref{Dm}) the dipolar energy is then given by

\begin{equation}\label{edip-ising}
    e_{dip}\sim\frac{\pi^2}{3} - \frac{8}{h} \sum_{m=1,3,\ldots}^{2h-1} \frac{1}{m} + \frac{4}{h}\sim \frac{\pi^2}{3} +4\frac{\psi(h)-\beta}{h},
\end{equation}

\noindent where $\beta\equiv \gamma_e +\ln\, 4-1$, $\gamma_e\approx0.577216$ is the Euler gamma constant and  $\psi(x)$ is the digamma function\cite{GrRi1994}. The energy per spin respect to the in--plane magnetized state is then given by

\begin{equation}
e_I =  -\kappa' + \frac{\pi^2}{3}+\frac{2\,\delta'-\,\beta}{h} -\frac{4\,\psi(h)}{h}
\label{energiaIsing}
\end{equation}

\noindent Minimizing Eq.(\ref{energiaIsing}) with respect to $h$ leads to the equation $\delta'/2-\beta= F(h)$, where $F(h)= \psi(h)-h\,\psi'(h)\sim \ln h -1$, thus recovering the known result $h\sim e^{\delta/2}$. Comparing the energies, we find that the HWP  has less energy than the Ising state for any value of $\eta$. Eq.(\ref{htanhYG}) shows that the stripe width variation in the Saturated state is determined by the change in the wall width as the anisotropy increases. Hence, $h$ will change until the wall width reaches the atomic limit, i.e. for $l_w=1$, where Eq.(\ref{htanhYG}) recovers the Ising behavior  $h\sim e^{\delta/2}$. Imposing the condition $l_w=1$ to Eq.(\ref{lwtanhYG}) we obtain the transition line between the Saturated and the Ising Stripes states:

\begin{equation}\label{etatanhYG}
    \eta  =\frac{1}{2}\, \delta -2 + \frac{\pi^2}{3} + 3g -2(c_2-c_1),
\end{equation}

\noindent which is also shown in Fig.\ref{pd2}, in complete agreement with the MC results.

\begin{figure}
\begin{center}
\includegraphics[scale=0.28,angle=-90]{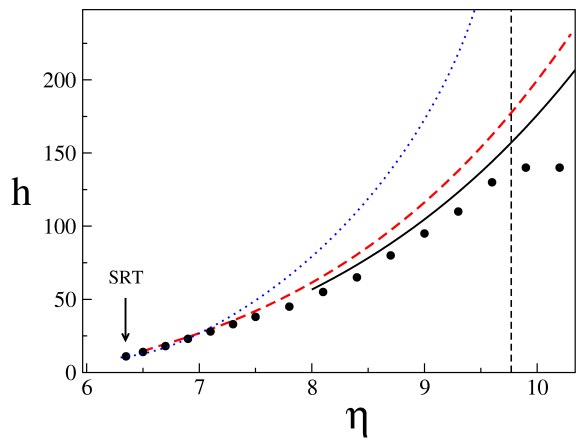}
\caption{\label{hvseta-delta10} (Color online) Comparison of the
  $T=0$  stripe width $h$ vs. $\eta$ obtained within the
  different methods for $\delta=10$. Symbols correspond to MC
  simulations. Full black  line corresponds to the asymptotic
  approximation for the HWP given by Eqs.(\ref{htanhYG}) and
  (\ref{lwtanhYG}). The red dashed line corresponds to the variational
  solution of Eqs.(\ref{anitanh}) and (\ref{eexchHWP2}) using the
  approximation (\ref{perfilTanh}) for the Fourier coefficients in the
  dipolar energy.  The blue dotted line corresponds to the continuous SWP. The vertical dashed line corresponds to the transition between Saturated and Ising Stripes states given by Eq.(\ref{etatanhYG}).}
\end{center}
\end{figure}

 In Fig.\ref{hvseta-delta10} we compare the
equilibrium  stripe width $h$ as a function of $\eta$ obtained within
the different approximations used in this work for $\delta=10$ and
with the MC simulations. Notice that the asymptotic approximation for
the HWP given by Eqs.(\ref{htanhYG}) and (\ref{lwtanhYG}) shows a
better agreement with the MC results than using the approximation (\ref{perfilTanh}) for the Fourier coefficients in the dipolar energy. This is because we adjusted the fitting parameter $f$ to optimize the agreement with the MC results at low values of $\delta$. The discrepancy between both (hyperbolic) approximations becomes negligible in the large $\delta$ limit.

\section{Discussion and conclusions}
\label{discu}

The main results of this work are summarized in Figs.\ref{pd1} and
\ref{pd2}, which display the complete zero temperature phase diagram
of the model defined by the Hamiltonian (\ref{hamiltoniano}).
Working upon reasonable assumptions for the ground states, like
perfectly straight modulations in one dimension and Bloch domain
walls, we analyzed minimum energy configurations combining a
variational analysis with Monte Carlo results. We found four
qualitatively different kinds of solutions: a planar ferromagnet for
small anisotropies, a Single Mode Canted Profile (characterized by  a sinusoidal variation of both component of the magnetization and varying wave length)  close to the SRT and two types of perpendicular striped states for large values of the anisotropy:  a saturated state in which the in-plane component is
restricted to the domain walls, and an Ising stripe state with sharp
walls for large anisotropies.

The SMCP and saturated states give valuable information on the
behavior of the stripe width (or the wave length in the SMCP case) as the anisotropy and exchange parameters
change. We find that stripe width
variation is directly associated to the presence of finite width
domain walls. For large enough values of the anisotropy $\eta$ the
ground state of the system is always an Ising Striped state, no matter
the value of the exchange coupling $\delta$. In those states domain
walls are sharp, the stripe width is completely independent of
$\eta$ and grows exponentially with the exchange coupling.

At the SRT the system passes through  canted-like states (mostly SMCP)as the
anisotropy increases, although the range of values of $\eta$ where the
canted angle is different from zero narrows as $\delta$ increases. For instance, the
exchange to dipolar coupling  ratio in fcc Fe based ultrathin films
can be roughly estimated to be $\delta \sim 100$ (considering a cubic
bilayer of Fe/Cu(100), where\cite{WuWoSc2004} the exchange coupling
$J_{Fe}\sim 30\, meV$, the lattice constant $d_{Fe} \sim 2 ML$
and\cite{DuArMa1996} $\mu_{Fe}\sim 3\,\mu_B$). For $\delta \sim 100$
the anisotropy interval for the canted-like states is approximately $\Delta \eta= \eta -\eta_{SRT} \approx 0.2$.

For  $\delta < 6$ the SMCP has the minimum energy in a rather  extended region of the phase parameters space, close to the SRT. The wave length (or ``stripe width'') of those states presents a strong variation with the anisotropy, directly correlated with an increasing canted angle. According to YG approximation, SMCP solutions are expected only close to the SRT. We found that the magnetization profile maintains the sinusoidal shape as the anisotropy increases. Above certain value of $\eta$ the wall profile changes to a hyperbolic tangent shape, while the magnetization inside the domains becomes fully saturated.

 For $\delta>6$ the ground state is given by the Saturated State, except very close to the SRT. A similar effect (i.e.  a crossover between a sinusoidal and a saturated magnetization profile) has been observed in room temperature grown fcc Fe/Cu(100) ultrathin films, as the temperature decreases from $T_c$, even though those systems do not present SRT\cite{ViSaPoPePo2008}.

 In the Saturated state, the stripe width increase with $\eta$  is
 directly related to the wall width decrease through the relation
 $h\sim e^{\delta/2l_w}$. The wall width in turn is determined by the
 competition between exchange and anisotropy. Once the anisotropy is
 large enough that the wall width reaches the atomic limit, $h$ growth
 stops. One may wonder whether a similar mechanism could be behind the stripe width variation with temperature, where a saturation is observed at low temperatures. Nevertheless, in
this case other effects, like extremely slow relaxation can be responsible for the
observed saturation.

Besides its direct application to real systems, knowing the ground state of this system for arbitrary values of the exchange coupling is of fundamental  importance to  have a correct interpretation of Monte Carlo simulation results. Being one of the most powerful tools to analyze these kind of systems at the present (specially at finite temperatures), it is basically limited by finite size restrictions, which implies relatively small values of $\delta$ (the characteristic length $h$ of the problem  grows exponentially with $\delta$ at low temperatures).

\section{Acknowledgments}

 We thank N. Saratz for advise about experimental results on Fe/Cu(001) ultrathin films.
This work was partially supported by grants from CONICET,
FONCyT grant PICT-2005 33305 , SeCyT-Universidad Nacional de C\'ordoba (Argentina),
CNPq and CAPES (Brazil), and ICTP grant NET-61 (Italy).

\appendix

\section{Zero Temperature Monte Carlo Technique for striped domain patterns}
\label{MCmethod}

In order to have an independent computation of the striped profiles which minimize the
energy, we implemented Monte Carlo simulation with a simulated
annealing protocol and Metropolis algorithm. To compare against the analytical solutions we  looked for minimum energy magnetization profiles among those characterized by periodic straight domains with Bloch walls. Hence, the problem is basically one dimensional and we could restrict the search to  one dimensional patterns over the $x$ direction fixing
$S_i^x=0$ $\forall i$ and imposing periodic boundary conditions (PBC) in both the $x$ and $y$ directions. In other words, we simulated a
 lattice with $L_x \times L_y$ with $L_y=1$ and PBC, which are implemented by means of the
Ewald sums technique.

The temperature was then decreased down to
very low temperatures at a constant rate $T(t)= T_0-r\,t$, where time is measured in Monte Carlo Steps, $T_0$ is the initial
temperature and $r$ is the cooling rate.  For all the range of parameters of this work, we choose $T_0=1$ and $r=10^{-4}$ and the simulation protocol was repeated $100$ times using different sequences of random numbers in order to minimize the possibility of trapping in local minima.
The results were independent of the initial spin configuration we choose at $T_0$.   For every set of values of $(\delta,\eta)$ we checked the results for
different  values of $L_x$ in order to avoid artificial frustration. We also performed some
 comparisons with MC results in a square $L_x=L_y$ lattice with PBC using the same annealing protocol and the results were
indistinguishable. This ansatz allowed us to obtain MC results for values of $\delta$ up to
$\delta=10$ (for which the maximum equilibrium value is $h=140$).


\end{document}